\documentclass[reprint,amsmath,amssymb,prl]{revtex4-1}
\usepackage{graphicx}
\usepackage{subfigure}
\graphicspath{ {Images/} }
\usepackage{bm}
\usepackage{hyperref}
\usepackage[mathlines]{lineno}
\usepackage[export]{adjustbox}
\usepackage{mathrsfs}
\usepackage{siunitx}
\usepackage{textcomp}
\usepackage[T1]{fontenc}

\begin{document}

\title{Single-Input Control of Multiple Fluid-Driven Elastic Actuators Via Interaction Between Bi-Stability and Viscosity} 
\author{$^{1}$Eran Ben-Haim}
\author{$^{2}$Lior Salem}
\author{$^{1,2}$Yizhar Or}
\author{$^{1,2}$Amir D. Gat}

 \affiliation{%
$^{1}$Faculty of Mechanical Engineering, Technion - Israel Institute of Technology, 32000 Haifa, Israel.
 }%
\affiliation{%
 $^{2}$Technion Autonomous Systems Program, Technion - Israel Institute of Technology ,3200003 Haifa, Israel
}%
\date{\today}


\begin{abstract}
A leading concept in soft robotics actuation, as well as in microfluidics applications such as valves in lab-on-a-chip devices, is applying pressurized flow in cavities embedded within elastic bodies. Generating complex deformation patterns typically requires control of several inputs, which greatly complicates the system's operation. In this work, we present a novel method for {\em single-input control} of a serial chain of {\em bi-stable} elastic chambers connected by thin tubes. Controlling a single flow rate at the chain's inlet, we induce an irreversible sequence of transitions that can reach any desired state combination of all bi-stable elements. Mathematical formulation and analysis of the system's dynamics reveal that these transitions are enabled thanks to bi-stability combined with pressure lag induced by viscous resistance. The results are demonstrated via numerical simulations combined with experiments for chains of up to 5 chambers, using water-diluted glycerol as the injected fluid. The proposed technique has a promising potential for development of sophisticated soft actuators with minimalistic control.
\end{abstract}


\maketitle

\section{\label{sec:level1}Introduction}
Soft robotics is a rapidly emerging concept where large continuous deformations and compliant interaction with external loads or environment are combined and harnessed for various applications of tactile manipulation, locomotion in unstructured environments, and bio-medical applications \cite{aguilar2016review,rus2015design}. A leading effective method of soft actuation is based on an elastic structure containing embedded network of cavities filled with fluid, while controlling pressures or flow rates at the network's inlets \cite{polygerinos2017soft,marchese2015recipe}.
Pressure control of multiple liquid-field elastic chambers is also commonly used to actuate onboard valves in the field of lab-on-a-chip devices, in which it is often required to dynamically change the geometry of internal micro-fluidic networks \cite{unger2000monolithic,thorsen2002microfluidic,desai2012design}. Generating and coordinating complex deformation patterns with such actuators (e.g. \cite{tolley2014resilient}) or micro-valves (e.g. \cite{mosadegh2010integrated}) typically requires control of several inputs, which greatly complicates the system's operation.

An additional feature exploited for efficient elastic actuation is using {\em bi-stability} of flexible elements for inducing rapid "jumps" between different stable equilibrium states due to excitation by minimal input \cite{rothemund2018soft}. Several works study the behavior of a serial chain of bi-stable elastic elements, where actuation can either be thermal \cite{chen2018autonomous,che2018viscoelastic}, electrical \cite{li2013giant,hines2016inflated} or tension forces \cite{Benichou2013,Puglisi2000,Cohen2014}. In the case of fluidic actuation, several works study variations of the well-known "two-balloon system", while others study networks of multiple connected chambers. Importantly, in many of these works, each element has its own control input for inducing transitions between its bi-stable states \cite{Fei2017,li2013giant,hines2016inflated}. Other works consider systems with a single control input \cite{Benichou2013,Puglisi2000,chen2018autonomous,che2018viscoelastic,Cohen2014} or no input \cite{Müller2004,Dreyer1982} (network with closed fluid domain), but do not allow for arbitrary control of transitions. The work \cite{overvelde2015amplifying} achieved a desired specific sequence of reversible state transitions of the bi-stable elements due to pre-planned mechanical tuning, but is also incapable of controlling and enforcing any desired cyclic and irreversible sequence of states without modifying the system's tuning. The work \cite{glozman2010self} has demonstrated crawling locomotion gait of a robot composed of serially connected bi-stable balloons. While this work has promising potential for biomedical applications, it was limited to achieving a specific gait of cyclic transition sequence and did not include systematic stability analysis of the system's states and possible transitions.

\begin{figure*}
    \centering
    \includegraphics[width=2.1\columnwidth, center]{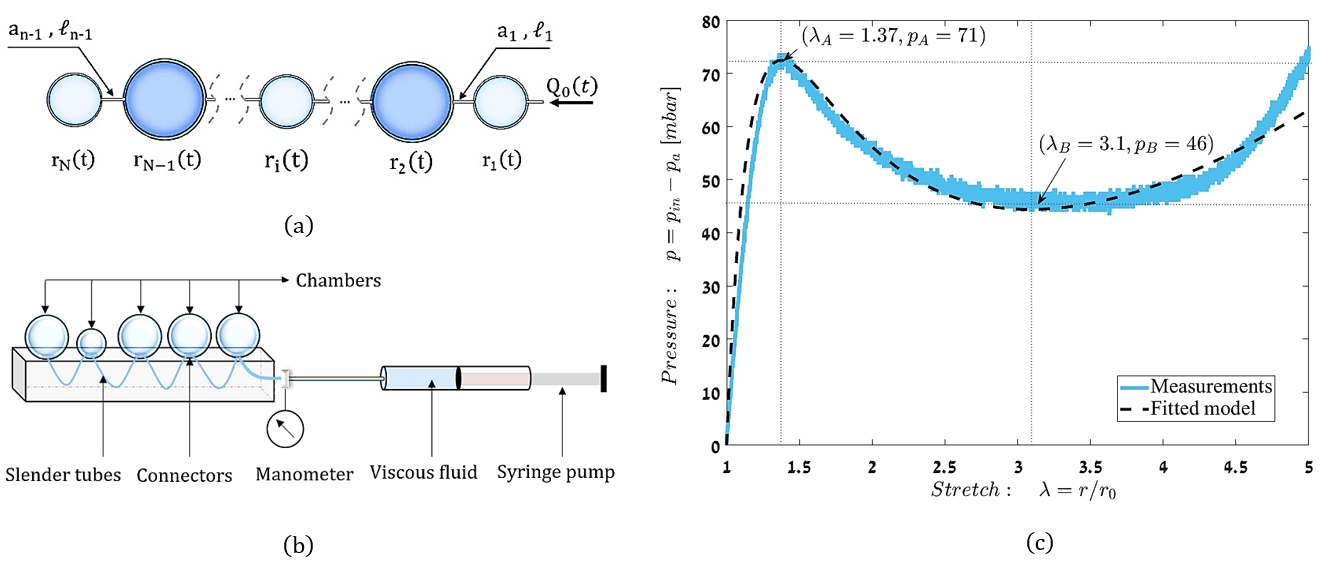}
    \caption{(a) Illustration of the system with notation. (b) Schematic picture of a serial chain of elastic chambers with a single inlet. (c) Characteristic stretch-pressure curve of a single elastic chamber obtained from experimental measurements (solid line) and the best-fit theoretical curve using Ogden's law \cite{ogden1972large}. 
    }
    \label{fig1}
\end{figure*}

In this work, we present a novel method for achieving arbitrary choices of state transition sequences in a serial chain of  bi-stable elastic chambers connected by thin tubes, using a single input of inlet flow rate. We present mathematical formulation of the system's dynamics and analyze its stability and transitions. The results are demonstrated via numerical simulations combined with laboratory experiments for chains of up to 5 chambers, using fluids of different viscosities. Schematic picture of our experimental setup and its illustration with notations are shown in Fig. \ref{fig1}(a) and \ref{fig1}(b), respectively.
\section{Formulation of elastic bi-stability}
We begin by introducing the main effect that enables state transitions in a pressurized thin elastic chamber, namely its \emph {bi-stability}, which induces multiple solutions of possible volumes for a given pressure. For an ideally spherical chamber with stress-free radius $r_0$ and shell thickness $d_0$, let $\lambda=r/r_0$ denote the {\em relative stretch} of the chamber, where $r$ is its varying radius. For a thin-shelled chamber made of incompressible hyper-elastic isotropic material such as rubber, finite elasticity theory dictates a known form of elastic strain energy density \cite{Treloar1975,holzapfel2002nonlinear,Müller2004,Vandermar2016}:
\begin{equation}  \label{eq:energy}
\psi(\lambda)=\sum_{k=1}^{K} {\frac{s_{k}}{\alpha_{k}}\big(2\lambda^{\alpha_{k}}+\lambda^{-2\alpha_{k}}-3\big)}
\end{equation}
The isotropic pressure in the shell can then be obtained from \eqref{eq:energy} as (cf. \cite{Beatty1987}):
\begin{equation} \label{eq:pressure}
p=\frac{d_{0}}{r_{0}}\frac{1}{\lambda^2}\cdot\frac{d\psi}{d\lambda}=2\frac{d_{0}}{r_{0}}\sum_{k=1}^{K} {s_{k}\big(\lambda^{\alpha_{k}-3}-\lambda^{-2\alpha_{k}-3}\big)}.
\end{equation}
Using the chamber's volume $V=\tfrac{4}{3}\pi r^3$, the pressure in \eqref{eq:pressure} can also be written in terms of $V$ as
\begin{equation} \label{eq:pressureV}
p:=F(V)=4 \pi d_0 r_0^2 \frac{d \psi}{d V}.
\end{equation}
Using a calibration experiment with a single chamber, Fig.\ref{fig1}(c) shows the measured stretch-pressure characteristic curve. The best fit model for this particular curve is Ogden's law \cite{ogden1972large}, which assumes powers of $\alpha_1=1.3,\; \alpha_2=5,\; \alpha_3=-2$ in \eqref{eq:pressure}. This curve appears in dashed line in Fig.  \ref{fig1}(c).  
The supplementary information (SI) contains details of the calibration experiment and the fitting process, as well as comparison with other known elasticity laws \cite{Treloar1975,Beatty1987,holzapfel2002nonlinear,Müller2004}. The characteristic curve $p(\lambda)$ in Fig.\ref{fig1}(c) has a local maximum point at $(\lambda_A,p_A)$ and a local minimum point at $(\lambda_B,p_B)$. For the intermediate range of pressures $p \in (p_A,p_B)$, there exist multiple solutions of stretch $\lambda$. We denote the stretch ranges $\lambda<\lambda_A$ and $\lambda>\lambda_B$ as "small" and "big" chambers, respectively, which are also represented by binary states $'0'$ and $'1'$. 
Analyzing the strain energy function $\psi(\lambda)$ in \eqref{eq:energy}, it can be proven that these upper and lower solution branches of $p(\lambda)$ are stable equilibria and satisfy $d^{2}\psi / d \lambda^{2} >0$. Conversely, the intermediate branch $\lambda \in (\lambda_A,\lambda_B)$ is an unstable solution satisfying $d^{2}\psi / d \lambda^{2} <0$. This is precisely the bi-stability phenomenon, which is exploited in this work for controlled transitions of between combinations of the chambers' binary states in a desired sequence.

 \begin{figure*}
    \centering
    \includegraphics[width=2.1\columnwidth, center]{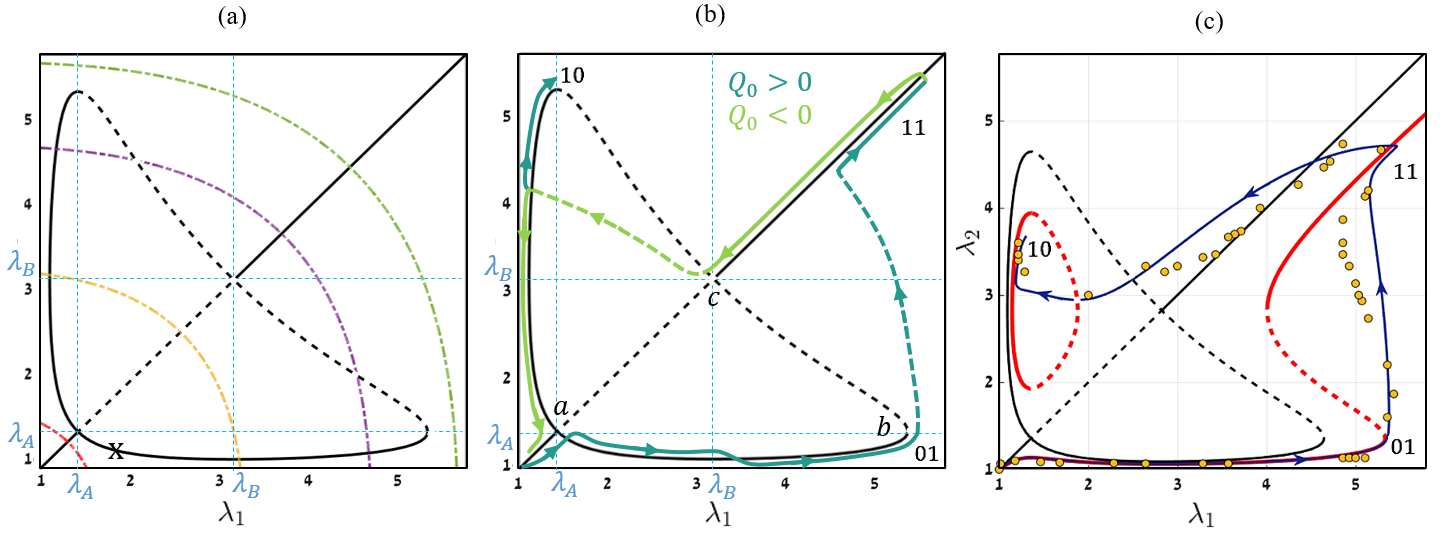}
    \caption{(a) Equilibrium curves of the two-chamber system in $(\lambda_1,\lambda_2)$ plane. Solid curves are stable branches and dashed curves are unstable ones. Dash-dotted cubic curves are constant total volume $\lambda_{1}^{3}+\lambda_{2}^{3}=Const$. (b) Solution trajectories of numerical simulation, overlaid on the branches of equilibrium curves. (c) Results of experimental measurements as circles in $(\lambda_1,\lambda_2)$-plane. Black lines are theoretical curves of equilibrium $p_1=p_2$ (solid - stable, dashed - unstable). Red thick lines are equilibrium curves of asymmetric chambers with $c=0.3$, showing good agreement with experimental measurements. }
    \label{fig2}
\end{figure*}


\section{Analysis of multi-chamber system}
We now consider a serial chain of $N$ identical chambers connected by thin circular tubes, as shown in Fig.\ref{fig1}(a). The system is filled by incompressible Newtonian fluid with density $\rho$ and viscosity $\mu$. We denote $Q_i$ as the mass flow rate at the tube connecting the chambers $i$ and $i+1$, where the only controlled input is the time-varying mass flow rate $Q_0(t)$ at the inlet of the first chamber. Each tube has radius of $a_i$ and length $l_i$. Denoting the volume of the $i^{th}$ chamber as $V_i=\tfrac{4}{3}\pi (r_0 \lambda_i)^3$, mass rate balance for each chamber gives:
\begin{equation} \label{eq:mass_rate}
\rho \frac{dV_{i}}{dt}=4\rho \pi r_0^3 \lambda_i^2 \frac{d \lambda_{i}}{dt} =  Q_{i-1}-Q_{i} \mbox{ for } i=1 \ldots N
\end{equation}
The fact that the chain has no outlet is represented by the end condition $Q_N=0$.
We assume steady and fully-developed axisymmetric flow in the tubes, with uniform pressure gradient. 
Neglecting end effects, this gives a linear relation between viscous flow rate and pressure difference at the $i^{th}$ tube as:
\begin{equation} \label{eq:Q}
Q_{i}=\frac{p_i-p_{i+1}}{R_{i}},
\quad\mbox{ where } R_i=\frac{8\mu l_i}{\pi \rho a_{i}^4},\; i\in [1,N-1].
\end{equation}

Finally, assuming quasistatic equilibrium and uniform pressure at each chamber implies that the fluid pressures $p_i$ at the $i^{th}$ chamber is equal to the pressure in the chamber's shell and satisfies the characteristic relation $p_i=F(V_i)$, where the function $F(V)$ is given in \eqref{eq:pressureV}. Substituting into \eqref{eq:Q} gives
\begin{equation} \label{eq:Q_lambdas}
Q_{i}=\frac{F(V_i)-F(V_{i+1})}{R_{i}},\quad i\in[1, N-1].
\end{equation}

Substituting equations \eqref{eq:Q_lambdas} into \eqref{eq:mass_rate} gives a nonlinear coupled system of $N$ first-order differential equations which govern the dynamic evolution of chambers' volumes $V_i(t)$ under the single input $Q_0(t)$. Equilibrium states of the system under zero input $Q_0=0$ for which the total volume is conserved impose that all chambers' pressures are equal $p_1=p_2 \ldots =p_N$. For a given total volume $V_{tot}=V_1 + \ldots + V_N$, this condition may have multiple solutions due to bi-stability of the stretch-pressure relation in \eqref{eq:pressure}, see Fig.\ref{fig1}(c). Assuming slow changes in the total volume, the system moves quasistatically along equilibrium solutions. However, any nonzero input $Q_{0}(t)$ still induces small deviations from equilibrium, and thus divergence from, or convergence to, solutions of static equilibrium is determined by their {\em dynamic stability}. Stability analysis of the system's equilibria is detailed in the SI. For the case of two identical chambers $N=2$, the condition for stability of an equilibrium state with volumes $(V_1^e,V_2^e)$ is given by:
\begin{equation} \label{eq:stable_two}
\left.\frac{dF}{d V}\right|_{V_1^e} +
\left.\frac{dF}{d V}\right|_{V_2^e} >0.
\end{equation}
For the case of three chambers $N=3$, the stability condition \eqref{eq:stable_two} still holds, and is augmented by the additional condition
\begin{equation} \label{eq:stable_three}
\left.\frac{dF}{d V}\right|_{V_1^e} \cdot \left.\frac{dF}{d V}\right|_{V_2^e} + \left.\frac{dF}{d V}\right|_{V_1^e} \cdot \left.\frac{dF}{d V}\right|_{V_3^e} + \left.\frac{dF}{d V}\right|_{V_2^e} \cdot \left.\frac{dF}{d V}\right|_{V_3^e} >0.
\end{equation}
Extension to general case of multiple chambers appears in the SI.

 \begin{figure*}
    \centering
    \includegraphics[width=2.1\columnwidth, center]{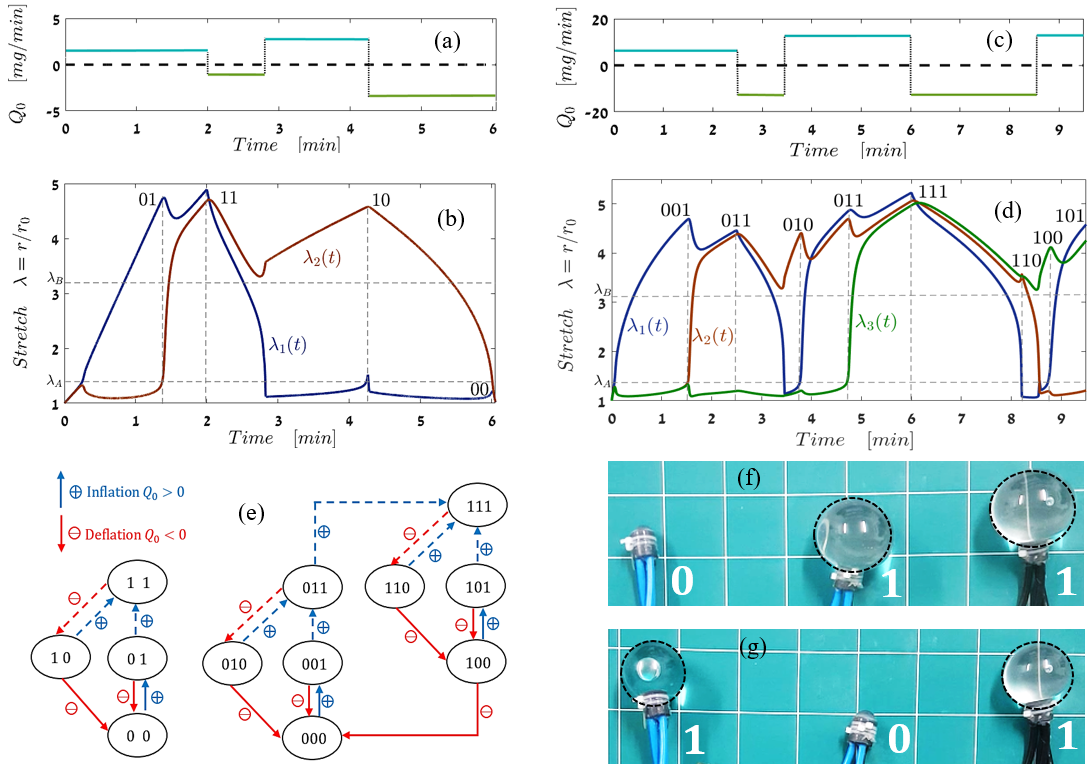}
    \caption{(a) Time plot of inlet flow $Q_0(t)$ for inflation and deflation in case of two chambers. (b) Time plots of chambers' stretches $\lambda_i(t)$ obtained by numerical integration of the nonlinear dynamical system with $N=2$ chambers. (c) Time plot of inlet flow $Q_0(t)$ in case of three chambers. (d) Time plots of chambers' stretches $\lambda_i(t)$ with $N=3$ chambers. (e) State transitions graph for systems of two and three chambers (left for $N=2$ and right for $N=3$), Dashed arrows represent transitions of "jumps" due to stability loss of previous solution. (f) and (g) are snapshots from experiment with $N=3$ chambers, showing example of two different binary states '011' and '101'.}
    \label{fig3}
\end{figure*}

In order to illustrate the concepts of multiple equilibrium solution branches and their dynamic stability, we first consider the two-chamber system $N=2$. A plot of the equilibrium solution branches in the plane of stretches $(\lambda_1,\lambda_2)$ under $Q_0=0$ 
is shown in Fig.\ref{fig2}. Using the stability condition \eqref{eq:stable_two}, solution branches of stable equilibria are marked by solid lines, while branches of unstable equilibria are marked by dashed lines. Importantly, stability condition for the system's equilibria is fundamentally different from stability of a single chamber based on its elastic potential $\psi$ in \eqref{eq:energy}. For example, the point marked by '$\times$' on Fig.\ref{fig2}(a) is a stable equilibrium state even though chamber 1 seems to be in an "unstable" solution branch $\lambda_1 \in (\lambda_A,\lambda_B)$.  The dash-dotted arcs in Fig.\ref{fig2}(a) denote curves of constant total volume $\lambda_1^3+\lambda_2^3=const$. When the system is initially placed out of equilibrium with $Q_0=0$, the solution moves along these curves and converges towards stable equilibrium branches. This plot can also provide an elegant explanation to the counter-intuitive behavior of the well-known two-balloon experiment \cite{weinhaus1978equilibrium,Müller2004}, see details in the SI.

Next, we consider a scenario where the system undergoes irreversible sequence of transitions between the chambers' combined states, while being controlled by a single input of flow rate $Q_0(t)$. The chosen input is piecewise constant. Fig.\ref{fig2}(b) shows the system's trajectory in $(\lambda_1,\lambda_2)$-plane, overlayed on the equilibrium curves. The plots show how the system goes through the irreversible sequence of states $00 \to 01 \to 11 \to 10 \to 00$, where the rightmost digit corresponds to the state of chamber 1. These state transitions are made possible by exploiting the following two key effects. First, when the state trajectory follows a stable branch and reaches a point where it becomes unstable, as in points 'a,b,c' in Fig.\ref{fig2}(b), the trajectory rapidly "jumps" and converges to a stable branch, moving very close to a cubic arc of constant total volume,  $\lambda_1^3+\lambda_2^3=const$. The second key effect, which dictates the "choice" of the direction to which the solution converges after loss of stability, is explained as a follows. Under small nonzero input, the sign of $Q_0$ combined with the current stretches $\lambda_i(t)$ dictate the sign of non-equilibrium pressure difference $ \Delta p= p_2-p_1 \neq 0$. This results in small deviation of the state trajectory from an equilibrium curve in a specific direction, as seen in Fig.\ref{fig2}(b). Thus, when the trajectory approaches bifurcation points 'a,c', deviations from the symmetry line $\lambda_1=\lambda_2$ imposed by the sign of $\Delta p$ dictate different "choices" of converging towards particular stable branches.
Fig.\ref{fig3}(a) shows time plot of the imput $Q_{0}(t)$ which represents a slow process of inflation followed by deflation. Fig.\ref{fig3}(b) shows time plots of the two stretches $\lambda_1(t)$, $\lambda_2(t)$, which are obtained via
numerical integration of the two-chamber system of differential equations.
\begin{figure*}
    \centering
    \includegraphics[width=1.95\columnwidth, center]{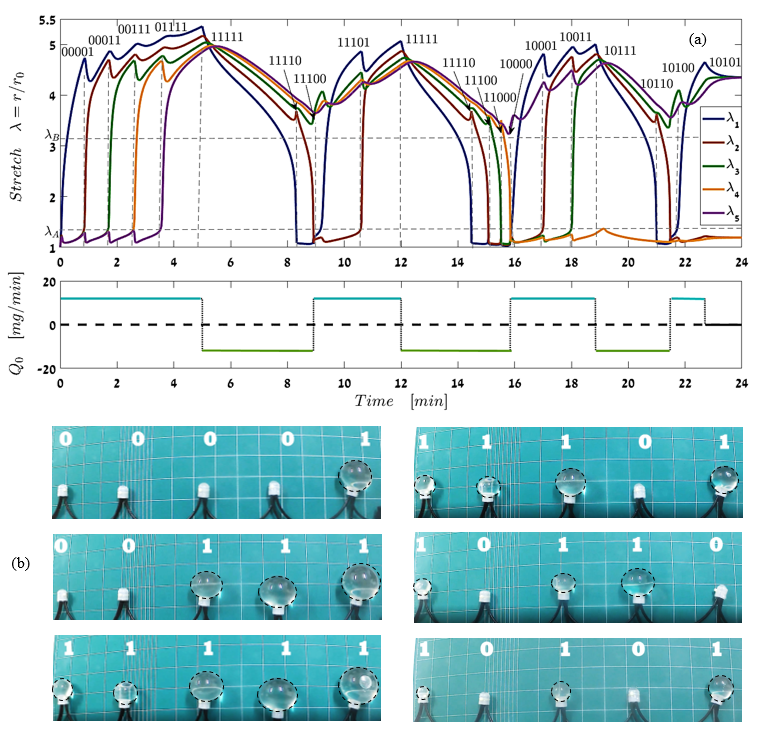}
    \caption{(a) Time plots of $Q_0(t)$ and $\lambda_i(t)$ for numerical simulation of the systems with $N=5$ chambers. Here we present 16 different binary states yielded by single input $Q_{0}(t)$ which plotted below this graph. (b) Snapshots of experiment with $N=5$ chambers, showing sequence of transitions between binary states.}
    \label{fig4}
\end{figure*}

These principles can be generalized to a serial chain of $N$ chambers, and induce a graph of possible transitions between binary states, achievable by using either inflation $Q_0>0$ or deflation $Q_0<0$. As an example, we present numerical simulations for chains of 3 and 5 chambers, and demonstrate generating a desired sequence of states. Time plots of the chosen input flow rate $Q_0(t)$ and the stretches $\lambda_i(t)$ are shown in Fig.\ref{fig3}(c) and $3$(d), and in Fig.\ref{fig4}(a).
Transition graphs for $N=2$ and $N=3$ are shown in Fig.\ref{fig3}(e), while graphs for $N=4,5$ appear in the SI.
Importantly, implementation of such ordered sequences of single-input state transitions does not depend on the particular details of the stretch-pressure characteristic curve in \eqref{eq:pressure} nor on specific pressure-flow rate relation \eqref{eq:mass_rate}. The only two necessary ingredients are the bi-stable form as shown in Fig.\ref{fig1}(b), combined with existence of non-equilibrium pressure deviation $\Delta p$ induced by viscous resistance of the tubes.

\section{Experiments}
Our experimental setup of $N=\{2,3,5\}$ elastic chambers connected serially via thin semi-flexible tubes. The chambers were made of industrial latex.
The hyper-elastic stretch-pressure relation of a single chamber has been measured and numerically fitted to Ogden's law, see Fig.\ref{fig1}(c). The inlet is connected to a flow controller \textit{'neMESYS XL 7000N'} of $\mathrm{Cetoni^\circledR}$, through a tube of length $l_0=0.5m$ and circular cross-section of radius $a_0=10mm$. The tubes connecting all other chambers have equal lengths of $l_i=0.4m$ and radius of $a_i=1.5mm$, for $i=1 \ldots N$. Most of the experiments were conducted with water-diluted glycerol of viscosity $\mu\approx 1.2 [Pa\cdot s]$ and density $\rho\approx 1.26 [g/cm^3]$. The high viscosity was exploited for increased lag and pressure differences $\Delta p$ between consecutive chambers, which helps in emphasizing the effect of selective irreversible state transitions. More details on the experiments and measurements, as well as video movie files of experiments appear in the SI. 
Fig.\ref{fig2}(c). shows state trajectory in the plane of $(\lambda_1,\lambda_2)$ obtained from measurements of experiment with $N=2$. The experiment shows irreversible state transition sequence of $00 \to 10 \to 11 \to 01$, as shown in the simulations above (Fig.\ref{fig2}(b) and \ref{fig3}(b)). The deviation between experimental measurements and theoretical equilibrium curves (overlaid on the plot in black) can be explained by imperfections of asymmetry between the two chambers. This can be captured in the theoretical model by assuming small differences in shell thickness $d_0$ and stress-free radius $r_0$ between the two chambers, which are manifested by replacing the elasticity relations in \eqref{eq:pressure} with $p_1=F(\lambda_1)$ and $p_2=(1+c)F(\lambda_2)$, where $c$ is a small factor representing this asymmetry. The red curves on the plot in Fig.\ref{fig2}(c) represent the asymmetric equilibrium branches $p_1=p_2$ under $c=0.3$. It can be seen that introducing this simple asymmetry factor into the model improves the quantitative agreement with the experimental measurements. Finally, we present results of experiments with $N=3$ and $N=5$ chambers. Snapshots of systems representative binary states of the system are shown in Fig.\ref{fig3}(f),$3$(g) and \ref{fig4}(b). A video movie of a sequence of transitions between 16 states for $N=5$ appears in the SI.

\section{Conclusion} In summary, we have presented a method for inducing prescribed sequences of state transitions in a serially connected chain of fluid-filled chambers using a single input of flow rate. The method exploits bi-stability of equilibrium states for hyper-elastic thin shells, as well as pressure differences induced by viscous effects. We presented theoretical analysis, numerical simulations, and experiments with inflatable balloons. This method enables overcoming the complexity of controlling each element using a separate input, and thus has a promising potential for creating minimal-control soft actuators. Although we were able to conduct also successful experiments of state transitions with air-inflated chambers, extending the theoretical formulation to ideal compressible gas, which complicates also the stability analysis, is relegated to future work. Finally, a challenging extension of this research includes design and operation of a legged soft robot composed of bi-stable inflatable cavities for creating locomotion gaits using single-input control.






\bibliography{references}

\begin{thebibliography}{28}%
\makeatletter
\providecommand \@ifxundefined [1]{%
 \@ifx{#1\undefined}
}%
\providecommand \@ifnum [1]{%
 \ifnum #1\expandafter \@firstoftwo
 \else \expandafter \@secondoftwo
 \fi
}%
\providecommand \@ifx [1]{%
 \ifx #1\expandafter \@firstoftwo
 \else \expandafter \@secondoftwo
 \fi
}%
\providecommand \natexlab [1]{#1}%
\providecommand \enquote  [1]{``#1''}%
\providecommand \bibnamefont  [1]{#1}%
\providecommand \bibfnamefont [1]{#1}%
\providecommand \citenamefont [1]{#1}%
\providecommand \href@noop [0]{\@secondoftwo}%
\providecommand \href [0]{\begingroup \@sanitize@url \@href}%
\providecommand \@href[1]{\@@startlink{#1}\@@href}%
\providecommand \@@href[1]{\endgroup#1\@@endlink}%
\providecommand \@sanitize@url [0]{\catcode `\\12\catcode `\$12\catcode
  `\&12\catcode `\#12\catcode `\^12\catcode `\_12\catcode `\%12\relax}%
\providecommand \@@startlink[1]{}%
\providecommand \@@endlink[0]{}%
\providecommand \url  [0]{\begingroup\@sanitize@url \@url }%
\providecommand \@url [1]{\endgroup\@href {#1}{\urlprefix }}%
\providecommand \urlprefix  [0]{URL }%
\providecommand \Eprint [0]{\href }%
\providecommand \doibase [0]{http://dx.doi.org/}%
\providecommand \selectlanguage [0]{\@gobble}%
\providecommand \bibinfo  [0]{\@secondoftwo}%
\providecommand \bibfield  [0]{\@secondoftwo}%
\providecommand \translation [1]{[#1]}%
\providecommand \BibitemOpen [0]{}%
\providecommand \bibitemStop [0]{}%
\providecommand \bibitemNoStop [0]{.\EOS\space}%
\providecommand \EOS [0]{\spacefactor3000\relax}%
\providecommand \BibitemShut  [1]{\csname bibitem#1\endcsname}%
\let\auto@bib@innerbib\@empty
\bibitem [{\citenamefont {Aguilar}\ \emph {et~al.}(2016)\citenamefont
  {Aguilar}, \citenamefont {Zhang}, \citenamefont {Qian}, \citenamefont
  {Kingsbury}, \citenamefont {McInroe}, \citenamefont {Mazouchova},
  \citenamefont {Li}, \citenamefont {Maladen}, \citenamefont {Gong},
  \citenamefont {Travers} \emph {et~al.}}]{aguilar2016review}%
  \BibitemOpen
  \bibfield  {author} {\bibinfo {author} {\bibfnamefont {J.}~\bibnamefont
  {Aguilar}}, \bibinfo {author} {\bibfnamefont {T.}~\bibnamefont {Zhang}},
  \bibinfo {author} {\bibfnamefont {F.}~\bibnamefont {Qian}}, \bibinfo {author}
  {\bibfnamefont {M.}~\bibnamefont {Kingsbury}}, \bibinfo {author}
  {\bibfnamefont {B.}~\bibnamefont {McInroe}}, \bibinfo {author} {\bibfnamefont
  {N.}~\bibnamefont {Mazouchova}}, \bibinfo {author} {\bibfnamefont
  {C.}~\bibnamefont {Li}}, \bibinfo {author} {\bibfnamefont {R.}~\bibnamefont
  {Maladen}}, \bibinfo {author} {\bibfnamefont {C.}~\bibnamefont {Gong}},
  \bibinfo {author} {\bibfnamefont {M.}~\bibnamefont {Travers}},  \emph
  {et~al.},\ }\href@noop {} {\bibfield  {journal} {\bibinfo  {journal} {Reports
  on Progress in Physics}\ }\textbf {\bibinfo {volume} {79}},\ \bibinfo {pages}
  {110001} (\bibinfo {year} {2016})}\BibitemShut {NoStop}%
\bibitem [{\citenamefont {Rus}\ and\ \citenamefont
  {Tolley}(2015)}]{rus2015design}%
  \BibitemOpen
  \bibfield  {author} {\bibinfo {author} {\bibfnamefont {D.}~\bibnamefont
  {Rus}}\ and\ \bibinfo {author} {\bibfnamefont {M.~T.}\ \bibnamefont
  {Tolley}},\ }\href@noop {} {\bibfield  {journal} {\bibinfo  {journal}
  {Nature}\ }\textbf {\bibinfo {volume} {521}},\ \bibinfo {pages} {467}
  (\bibinfo {year} {2015})}\BibitemShut {NoStop}%
\bibitem [{\citenamefont {Polygerinos}\ \emph {et~al.}(2017)\citenamefont
  {Polygerinos}, \citenamefont {Correll}, \citenamefont {Morin}, \citenamefont
  {Mosadegh}, \citenamefont {Onal}, \citenamefont {Petersen}, \citenamefont
  {Cianchetti}, \citenamefont {Tolley},\ and\ \citenamefont
  {Shepherd}}]{polygerinos2017soft}%
  \BibitemOpen
  \bibfield  {author} {\bibinfo {author} {\bibfnamefont {P.}~\bibnamefont
  {Polygerinos}}, \bibinfo {author} {\bibfnamefont {N.}~\bibnamefont
  {Correll}}, \bibinfo {author} {\bibfnamefont {S.~A.}\ \bibnamefont {Morin}},
  \bibinfo {author} {\bibfnamefont {B.}~\bibnamefont {Mosadegh}}, \bibinfo
  {author} {\bibfnamefont {C.~D.}\ \bibnamefont {Onal}}, \bibinfo {author}
  {\bibfnamefont {K.}~\bibnamefont {Petersen}}, \bibinfo {author}
  {\bibfnamefont {M.}~\bibnamefont {Cianchetti}}, \bibinfo {author}
  {\bibfnamefont {M.~T.}\ \bibnamefont {Tolley}}, \ and\ \bibinfo {author}
  {\bibfnamefont {R.~F.}\ \bibnamefont {Shepherd}},\ }\href@noop {} {\bibfield
  {journal} {\bibinfo  {journal} {Advanced Engineering Materials}\ }\textbf
  {\bibinfo {volume} {19}},\ \bibinfo {pages} {1700016} (\bibinfo {year}
  {2017})}\BibitemShut {NoStop}%
\bibitem [{\citenamefont {Marchese}\ \emph {et~al.}(2015)\citenamefont
  {Marchese}, \citenamefont {Katzschmann},\ and\ \citenamefont
  {Rus}}]{marchese2015recipe}%
  \BibitemOpen
  \bibfield  {author} {\bibinfo {author} {\bibfnamefont {A.~D.}\ \bibnamefont
  {Marchese}}, \bibinfo {author} {\bibfnamefont {R.~K.}\ \bibnamefont
  {Katzschmann}}, \ and\ \bibinfo {author} {\bibfnamefont {D.}~\bibnamefont
  {Rus}},\ }\href@noop {} {\bibfield  {journal} {\bibinfo  {journal} {Soft
  Robotics}\ }\textbf {\bibinfo {volume} {2}},\ \bibinfo {pages} {7} (\bibinfo
  {year} {2015})}\BibitemShut {NoStop}%
\bibitem [{\citenamefont {Unger}\ \emph {et~al.}(2000)\citenamefont {Unger},
  \citenamefont {Chou}, \citenamefont {Thorsen}, \citenamefont {Scherer},\ and\
  \citenamefont {Quake}}]{unger2000monolithic}%
  \BibitemOpen
  \bibfield  {author} {\bibinfo {author} {\bibfnamefont {M.~A.}\ \bibnamefont
  {Unger}}, \bibinfo {author} {\bibfnamefont {H.-P.}\ \bibnamefont {Chou}},
  \bibinfo {author} {\bibfnamefont {T.}~\bibnamefont {Thorsen}}, \bibinfo
  {author} {\bibfnamefont {A.}~\bibnamefont {Scherer}}, \ and\ \bibinfo
  {author} {\bibfnamefont {S.~R.}\ \bibnamefont {Quake}},\ }\href@noop {}
  {\bibfield  {journal} {\bibinfo  {journal} {Science}\ }\textbf {\bibinfo
  {volume} {288}},\ \bibinfo {pages} {113} (\bibinfo {year}
  {2000})}\BibitemShut {NoStop}%
\bibitem [{\citenamefont {Thorsen}\ \emph {et~al.}(2002)\citenamefont
  {Thorsen}, \citenamefont {Maerkl},\ and\ \citenamefont
  {Quake}}]{thorsen2002microfluidic}%
  \BibitemOpen
  \bibfield  {author} {\bibinfo {author} {\bibfnamefont {T.}~\bibnamefont
  {Thorsen}}, \bibinfo {author} {\bibfnamefont {S.~J.}\ \bibnamefont {Maerkl}},
  \ and\ \bibinfo {author} {\bibfnamefont {S.~R.}\ \bibnamefont {Quake}},\
  }\href@noop {} {\bibfield  {journal} {\bibinfo  {journal} {Science}\ }\textbf
  {\bibinfo {volume} {298}},\ \bibinfo {pages} {580} (\bibinfo {year}
  {2002})}\BibitemShut {NoStop}%
\bibitem [{\citenamefont {Desai}\ \emph {et~al.}(2012)\citenamefont {Desai},
  \citenamefont {Tice}, \citenamefont {Apblett},\ and\ \citenamefont
  {Kenis}}]{desai2012design}%
  \BibitemOpen
  \bibfield  {author} {\bibinfo {author} {\bibfnamefont {A.~V.}\ \bibnamefont
  {Desai}}, \bibinfo {author} {\bibfnamefont {J.~D.}\ \bibnamefont {Tice}},
  \bibinfo {author} {\bibfnamefont {C.~A.}\ \bibnamefont {Apblett}}, \ and\
  \bibinfo {author} {\bibfnamefont {P.~J.}\ \bibnamefont {Kenis}},\ }\href@noop
  {} {\bibfield  {journal} {\bibinfo  {journal} {Lab on a Chip}\ }\textbf
  {\bibinfo {volume} {12}},\ \bibinfo {pages} {1078} (\bibinfo {year}
  {2012})}\BibitemShut {NoStop}%
\bibitem [{\citenamefont {Tolley}\ \emph {et~al.}(2014)\citenamefont {Tolley},
  \citenamefont {Shepherd}, \citenamefont {Mosadegh}, \citenamefont {Galloway},
  \citenamefont {Wehner}, \citenamefont {Karpelson}, \citenamefont {Wood},\
  and\ \citenamefont {Whitesides}}]{tolley2014resilient}%
  \BibitemOpen
  \bibfield  {author} {\bibinfo {author} {\bibfnamefont {M.~T.}\ \bibnamefont
  {Tolley}}, \bibinfo {author} {\bibfnamefont {R.~F.}\ \bibnamefont
  {Shepherd}}, \bibinfo {author} {\bibfnamefont {B.}~\bibnamefont {Mosadegh}},
  \bibinfo {author} {\bibfnamefont {K.~C.}\ \bibnamefont {Galloway}}, \bibinfo
  {author} {\bibfnamefont {M.}~\bibnamefont {Wehner}}, \bibinfo {author}
  {\bibfnamefont {M.}~\bibnamefont {Karpelson}}, \bibinfo {author}
  {\bibfnamefont {R.~J.}\ \bibnamefont {Wood}}, \ and\ \bibinfo {author}
  {\bibfnamefont {G.~M.}\ \bibnamefont {Whitesides}},\ }\href@noop {}
  {\bibfield  {journal} {\bibinfo  {journal} {Soft robotics}\ }\textbf
  {\bibinfo {volume} {1}},\ \bibinfo {pages} {213} (\bibinfo {year}
  {2014})}\BibitemShut {NoStop}%
\bibitem [{\citenamefont {Mosadegh}\ \emph {et~al.}(2010)\citenamefont
  {Mosadegh}, \citenamefont {Kuo}, \citenamefont {Tung}, \citenamefont
  {Torisawa}, \citenamefont {Bersano-Begey}, \citenamefont {Tavana},\ and\
  \citenamefont {Takayama}}]{mosadegh2010integrated}%
  \BibitemOpen
  \bibfield  {author} {\bibinfo {author} {\bibfnamefont {B.}~\bibnamefont
  {Mosadegh}}, \bibinfo {author} {\bibfnamefont {C.-H.}\ \bibnamefont {Kuo}},
  \bibinfo {author} {\bibfnamefont {Y.-C.}\ \bibnamefont {Tung}}, \bibinfo
  {author} {\bibfnamefont {Y.-s.}\ \bibnamefont {Torisawa}}, \bibinfo {author}
  {\bibfnamefont {T.}~\bibnamefont {Bersano-Begey}}, \bibinfo {author}
  {\bibfnamefont {H.}~\bibnamefont {Tavana}}, \ and\ \bibinfo {author}
  {\bibfnamefont {S.}~\bibnamefont {Takayama}},\ }\href@noop {} {\bibfield
  {journal} {\bibinfo  {journal} {Nature physics}\ }\textbf {\bibinfo {volume}
  {6}},\ \bibinfo {pages} {433} (\bibinfo {year} {2010})}\BibitemShut {NoStop}%
\bibitem [{\citenamefont {Rothemund}\ \emph {et~al.}(2018)\citenamefont
  {Rothemund}, \citenamefont {Ainla}, \citenamefont {Belding}, \citenamefont
  {Preston}, \citenamefont {Kurihara}, \citenamefont {Suo},\ and\ \citenamefont
  {Whitesides}}]{rothemund2018soft}%
  \BibitemOpen
  \bibfield  {author} {\bibinfo {author} {\bibfnamefont {P.}~\bibnamefont
  {Rothemund}}, \bibinfo {author} {\bibfnamefont {A.}~\bibnamefont {Ainla}},
  \bibinfo {author} {\bibfnamefont {L.}~\bibnamefont {Belding}}, \bibinfo
  {author} {\bibfnamefont {D.~J.}\ \bibnamefont {Preston}}, \bibinfo {author}
  {\bibfnamefont {S.}~\bibnamefont {Kurihara}}, \bibinfo {author}
  {\bibfnamefont {Z.}~\bibnamefont {Suo}}, \ and\ \bibinfo {author}
  {\bibfnamefont {G.~M.}\ \bibnamefont {Whitesides}},\ }\href@noop {}
  {\bibfield  {journal} {\bibinfo  {journal} {Science Robotics}\ }\textbf
  {\bibinfo {volume} {3}},\ \bibinfo {pages} {eaar7986} (\bibinfo {year}
  {2018})}\BibitemShut {NoStop}%
\bibitem [{\citenamefont {Chen}\ and\ \citenamefont
  {Shea}(2018)}]{chen2018autonomous}%
  \BibitemOpen
  \bibfield  {author} {\bibinfo {author} {\bibfnamefont {T.}~\bibnamefont
  {Chen}}\ and\ \bibinfo {author} {\bibfnamefont {K.}~\bibnamefont {Shea}},\
  }\href@noop {} {\bibfield  {journal} {\bibinfo  {journal} {3D Printing and
  Additive Manufacturing}\ } (\bibinfo {year} {2018})}\BibitemShut {NoStop}%
\bibitem [{\citenamefont {Che}\ \emph {et~al.}(2018)\citenamefont {Che},
  \citenamefont {Yuan}, \citenamefont {Qi},\ and\ \citenamefont
  {Meaud}}]{che2018viscoelastic}%
  \BibitemOpen
  \bibfield  {author} {\bibinfo {author} {\bibfnamefont {K.}~\bibnamefont
  {Che}}, \bibinfo {author} {\bibfnamefont {C.}~\bibnamefont {Yuan}}, \bibinfo
  {author} {\bibfnamefont {H.~J.}\ \bibnamefont {Qi}}, \ and\ \bibinfo {author}
  {\bibfnamefont {J.}~\bibnamefont {Meaud}},\ }\href@noop {} {\bibfield
  {journal} {\bibinfo  {journal} {Soft matter}\ }\textbf {\bibinfo {volume}
  {14}},\ \bibinfo {pages} {2492} (\bibinfo {year} {2018})}\BibitemShut
  {NoStop}%
\bibitem [{\citenamefont {Li}\ \emph {et~al.}(2013)\citenamefont {Li},
  \citenamefont {Keplinger}, \citenamefont {Baumgartner}, \citenamefont
  {Bauer}, \citenamefont {Yang},\ and\ \citenamefont {Suo}}]{li2013giant}%
  \BibitemOpen
  \bibfield  {author} {\bibinfo {author} {\bibfnamefont {T.}~\bibnamefont
  {Li}}, \bibinfo {author} {\bibfnamefont {C.}~\bibnamefont {Keplinger}},
  \bibinfo {author} {\bibfnamefont {R.}~\bibnamefont {Baumgartner}}, \bibinfo
  {author} {\bibfnamefont {S.}~\bibnamefont {Bauer}}, \bibinfo {author}
  {\bibfnamefont {W.}~\bibnamefont {Yang}}, \ and\ \bibinfo {author}
  {\bibfnamefont {Z.}~\bibnamefont {Suo}},\ }\href@noop {} {\bibfield
  {journal} {\bibinfo  {journal} {Journal of the Mechanics and Physics of
  Solids}\ }\textbf {\bibinfo {volume} {61}},\ \bibinfo {pages} {611} (\bibinfo
  {year} {2013})}\BibitemShut {NoStop}%
\bibitem [{\citenamefont {Hines}\ \emph {et~al.}(2016)\citenamefont {Hines},
  \citenamefont {Petersen},\ and\ \citenamefont {Sitti}}]{hines2016inflated}%
  \BibitemOpen
  \bibfield  {author} {\bibinfo {author} {\bibfnamefont {L.}~\bibnamefont
  {Hines}}, \bibinfo {author} {\bibfnamefont {K.}~\bibnamefont {Petersen}}, \
  and\ \bibinfo {author} {\bibfnamefont {M.}~\bibnamefont {Sitti}},\
  }\href@noop {} {\bibfield  {journal} {\bibinfo  {journal} {Advanced
  Materials}\ }\textbf {\bibinfo {volume} {28}},\ \bibinfo {pages} {3690}
  (\bibinfo {year} {2016})}\BibitemShut {NoStop}%
\bibitem [{\citenamefont {Benichou}\ \emph {et~al.}(2013)\citenamefont
  {Benichou}, \citenamefont {Faran}, \citenamefont {Shilo},\ and\ \citenamefont
  {Givli}}]{Benichou2013}%
  \BibitemOpen
  \bibfield  {author} {\bibinfo {author} {\bibfnamefont {I.}~\bibnamefont
  {Benichou}}, \bibinfo {author} {\bibfnamefont {E.}~\bibnamefont {Faran}},
  \bibinfo {author} {\bibfnamefont {D.}~\bibnamefont {Shilo}}, \ and\ \bibinfo
  {author} {\bibfnamefont {S.}~\bibnamefont {Givli}},\ }\href@noop {}
  {\bibfield  {journal} {\bibinfo  {journal} {Applied Physics Letters}\
  }\textbf {\bibinfo {volume} {102}},\ \bibinfo {pages} {011912} (\bibinfo
  {year} {2013})}\BibitemShut {NoStop}%
\bibitem [{\citenamefont {Puglisi}\ and\ \citenamefont
  {Truskinovsky}(2000)}]{Puglisi2000}%
  \BibitemOpen
  \bibfield  {author} {\bibinfo {author} {\bibfnamefont {G.}~\bibnamefont
  {Puglisi}}\ and\ \bibinfo {author} {\bibfnamefont {L.}~\bibnamefont
  {Truskinovsky}},\ }\href@noop {} {\bibfield  {journal} {\bibinfo  {journal}
  {Journal of the Mechanics and Physics of Solids}\ }\textbf {\bibinfo {volume}
  {48}},\ \bibinfo {pages} {1} (\bibinfo {year} {2000})}\BibitemShut {NoStop}%
\bibitem [{\citenamefont {Cohen}\ and\ \citenamefont
  {Givli}(2014)}]{Cohen2014}%
  \BibitemOpen
  \bibfield  {author} {\bibinfo {author} {\bibfnamefont {T.}~\bibnamefont
  {Cohen}}\ and\ \bibinfo {author} {\bibfnamefont {S.}~\bibnamefont {Givli}},\
  }\href@noop {} {\bibfield  {journal} {\bibinfo  {journal} {Journal of the
  Mechanics and Physics of Solids}\ }\textbf {\bibinfo {volume} {64}},\
  \bibinfo {pages} {426} (\bibinfo {year} {2014})}\BibitemShut {NoStop}%
\bibitem [{\citenamefont {Fei}\ and\ \citenamefont {Pang}(2017)}]{Fei2017}%
  \BibitemOpen
  \bibfield  {author} {\bibinfo {author} {\bibfnamefont {Y.}~\bibnamefont
  {Fei}}\ and\ \bibinfo {author} {\bibfnamefont {W.}~\bibnamefont {Pang}},\
  }\href@noop {} {\bibfield  {journal} {\bibinfo  {journal} {Nonlinear
  Dynamics}\ }\textbf {\bibinfo {volume} {88}},\ \bibinfo {pages} {883}
  (\bibinfo {year} {2017})}\BibitemShut {NoStop}%
\bibitem [{\citenamefont {Müller}\ and\ \citenamefont
  {Strehlow}(2004)}]{Müller2004}%
  \BibitemOpen
  \bibfield  {author} {\bibinfo {author} {\bibfnamefont {I.}~\bibnamefont
  {Müller}}\ and\ \bibinfo {author} {\bibfnamefont {P.}~\bibnamefont
  {Strehlow}},\ }\href@noop {} {\emph {\bibinfo {title} {Rubber and rubber
  balloons: paradigms of thermodynamics}}},\ Vol.\ \bibinfo {volume} {637}\
  (\bibinfo  {publisher} {Springer Science \& Business Media},\ \bibinfo {year}
  {2004})\BibitemShut {NoStop}%
\bibitem [{\citenamefont {Dreyer}\ \emph {et~al.}(1982)\citenamefont {Dreyer},
  \citenamefont {Müller},\ and\ \citenamefont {Strehlow}}]{Dreyer1982}%
  \BibitemOpen
  \bibfield  {author} {\bibinfo {author} {\bibfnamefont {W.}~\bibnamefont
  {Dreyer}}, \bibinfo {author} {\bibfnamefont {I.}~\bibnamefont {Müller}}, \
  and\ \bibinfo {author} {\bibfnamefont {P.}~\bibnamefont {Strehlow}},\
  }\href@noop {} {\bibfield  {journal} {\bibinfo  {journal} {The Quarterly
  Journal of Mechanics and Applied Mathematics}\ }\textbf {\bibinfo {volume}
  {35}},\ \bibinfo {pages} {419} (\bibinfo {year} {1982})}\BibitemShut
  {NoStop}%
\bibitem [{\citenamefont {Overvelde}\ \emph {et~al.}(2015)\citenamefont
  {Overvelde}, \citenamefont {Kloek}, \citenamefont {D’haen},\ and\
  \citenamefont {Bertoldi}}]{overvelde2015amplifying}%
  \BibitemOpen
  \bibfield  {author} {\bibinfo {author} {\bibfnamefont {J.~T.}\ \bibnamefont
  {Overvelde}}, \bibinfo {author} {\bibfnamefont {T.}~\bibnamefont {Kloek}},
  \bibinfo {author} {\bibfnamefont {J.~J.}\ \bibnamefont {D’haen}}, \ and\
  \bibinfo {author} {\bibfnamefont {K.}~\bibnamefont {Bertoldi}},\ }\href@noop
  {} {\bibfield  {journal} {\bibinfo  {journal} {Proceedings of the National
  Academy of Sciences}\ }\textbf {\bibinfo {volume} {112}},\ \bibinfo {pages}
  {10863} (\bibinfo {year} {2015})}\BibitemShut {NoStop}%
\bibitem [{\citenamefont {Glozman}\ \emph {et~al.}(2010)\citenamefont
  {Glozman}, \citenamefont {Hassidov}, \citenamefont {Senesh},\ and\
  \citenamefont {Shoham}}]{glozman2010self}%
  \BibitemOpen
  \bibfield  {author} {\bibinfo {author} {\bibfnamefont {D.}~\bibnamefont
  {Glozman}}, \bibinfo {author} {\bibfnamefont {N.}~\bibnamefont {Hassidov}},
  \bibinfo {author} {\bibfnamefont {M.}~\bibnamefont {Senesh}}, \ and\ \bibinfo
  {author} {\bibfnamefont {M.}~\bibnamefont {Shoham}},\ }\href@noop {}
  {\bibfield  {journal} {\bibinfo  {journal} {IEEE Transactions on Biomedical
  Engineering}\ }\textbf {\bibinfo {volume} {57}},\ \bibinfo {pages} {1264}
  (\bibinfo {year} {2010})}\BibitemShut {NoStop}%
\bibitem [{\citenamefont {Ogden}(1972)}]{ogden1972large}%
  \BibitemOpen
  \bibfield  {author} {\bibinfo {author} {\bibfnamefont {R.~W.}\ \bibnamefont
  {Ogden}},\ }\href@noop {} {\bibfield  {journal} {\bibinfo  {journal} {Proc.
  R. Soc. Lond. A}\ }\textbf {\bibinfo {volume} {326}},\ \bibinfo {pages} {565}
  (\bibinfo {year} {1972})}\BibitemShut {NoStop}%
\bibitem [{\citenamefont {Treloar}(1975)}]{Treloar1975}%
  \BibitemOpen
  \bibfield  {author} {\bibinfo {author} {\bibfnamefont {L.~R.~G.}\
  \bibnamefont {Treloar}},\ }\href@noop {} {\emph {\bibinfo {title} {The
  physics of rubber elasticity}}}\ (\bibinfo  {publisher} {Oxford University
  Press, USA},\ \bibinfo {year} {1975})\BibitemShut {NoStop}%
\bibitem [{\citenamefont {Holzapfel}(2002)}]{holzapfel2002nonlinear}%
  \BibitemOpen
  \bibfield  {author} {\bibinfo {author} {\bibfnamefont {G.~A.}\ \bibnamefont
  {Holzapfel}},\ }\href@noop {} {\bibfield  {journal} {\bibinfo  {journal}
  {Meccanica}\ }\textbf {\bibinfo {volume} {37}},\ \bibinfo {pages} {489}
  (\bibinfo {year} {2002})}\BibitemShut {NoStop}%
\bibitem [{\citenamefont {Vandermarlière}(2016)}]{Vandermar2016}%
  \BibitemOpen
  \bibfield  {author} {\bibinfo {author} {\bibfnamefont {J.}~\bibnamefont
  {Vandermarlière}},\ }\href@noop {} {\bibfield  {journal} {\bibinfo
  {journal} {The Physics Teacher}\ }\textbf {\bibinfo {volume} {54}},\ \bibinfo
  {pages} {566} (\bibinfo {year} {2016})}\BibitemShut {NoStop}%
\bibitem [{\citenamefont {Beatty}(1987)}]{Beatty1987}%
  \BibitemOpen
  \bibfield  {author} {\bibinfo {author} {\bibfnamefont {M.~F.}\ \bibnamefont
  {Beatty}},\ }\href@noop {} {\bibfield  {journal} {\bibinfo  {journal}
  {Applied Mechanics Reviews}\ }\textbf {\bibinfo {volume} {40}},\ \bibinfo
  {pages} {1699} (\bibinfo {year} {1987})}\BibitemShut {NoStop}%
\bibitem [{\citenamefont {Weinhaus}\ and\ \citenamefont
  {Barker}(1978)}]{weinhaus1978equilibrium}%
  \BibitemOpen
  \bibfield  {author} {\bibinfo {author} {\bibfnamefont {F.}~\bibnamefont
  {Weinhaus}}\ and\ \bibinfo {author} {\bibfnamefont {W.}~\bibnamefont
  {Barker}},\ }\href@noop {} {\bibfield  {journal} {\bibinfo  {journal}
  {American Journal of Physics}\ }\textbf {\bibinfo {volume} {46}},\ \bibinfo
  {pages} {978} (\bibinfo {year} {1978})}\BibitemShut {NoStop}%
\end{thebibliography}%
\end{document}